\documentclass[%
 reprint,
 amsmath,amssymb,
 aps,
aas_macros]{revtex4-2}

\usepackage{graphicx}
\usepackage{dcolumn}
\usepackage{bm}
\usepackage{ulem}[normalem]
\usepackage{xcolor}
\usepackage{amssymb}
\usepackage{aas_macros}
\usepackage{hyperref}

\begin{document}




\title{On the maximal velocity of colliding galaxies}






\author{Anastasiia M. Osipova}
\affiliation{Astro Space Center of P.N. Lebedev Physical Institute, Moscow, Russia}
\affiliation{HSE University, 101000 Moscow, Russia}

\author{Sergey V. Pilipenko}
\email{spilipenko@asc.rssi.ru}
\affiliation{Astro Space Center of P.N. Lebedev Physical Institute, Moscow, Russia}
\affiliation{HSE University, 101000 Moscow, Russia}

\date{\today}

\begin{abstract}
In the current galaxy formation paradigm, collisions play a crucial role. A fraction of galaxy collisions results in flyby events, and a galaxy that has passed through another galaxy is called a backsplash galaxy. Such flyby events are of particular interest for explaining the quenching of isolated galaxies. One signature of backsplash galaxies is that they have high velocities relative to their environment, since they do not move in the same flow as surrounding galaxies. This feature can be studied in simulations, but it is also useful to have a theory that can predict the velocities of backsplash galaxies. In this paper, we develop such a theory based on the Zel'dovich approximation and use it to determine the maximal expected velocity of a backsplash galaxy in a given volume.
\end{abstract}

\maketitle






\section{Introduction}
\label{sec:introduction}
It is well known that galaxies experience collisions with other galaxies. Most of them result in mergers, but simulations show that, at least for the $\Lambda$CDM cosmology, some fraction of collisions may result in flybys. In a pair of galaxies that collided in the past, the smaller one is sometimes called a backsplash and the larger one is called a target. Studies of simulations \cite{Ludlow09,Teyssier12,GarrisonKimmel14,Bakels21,Green21} have found that about 10\% of the Local Group field haloes are backsplashers. In larger volumes, depending on halo mass, backsplashers constitute from $\sim2$\% for $10^{11}$~M$_\odot$ haloes to $\sim10$\% for $10^{8}$~M$_\odot$ haloes \cite{Diemer_2021,Osipova23}.

Such collisions, which result in the appearance of backsplash galaxies, could be very interesting for theories of galaxy evolution. One of the important processes in this evolution is quenching of star formation. It can be caused by a number of complex processes, which are usually divided into internal and external. It is widely believed that in the more massive systems internal processes dominate the quenching, while in dwarfs it is governed by the environment. For a long time this has been supported by the fact that almost all field dwarfs are star-forming. Only few field quiescent dwarfs in the mass range $10^7 < [M_\star/M_\odot] < 10^9$ were known, most of them are close to the Local Group \cite{Lavery92,Whiting99,Karachentsev01,McConnachie08,Makarov12,Karachentsev15,Makarova17,Polzin21,Sand22}.

A recent paper \cite{Bidaran_2025} claims that a sample of more than 20 quenched isolated dwarf galaxies ($M_\star<10^{9.5}$~M$_\odot$) in voids could indicate that internal mechanisms can also drive quenching in dwarfs. The authors of \cite{Bidaran_2025} used a criterion of isolation that there are no other galaxies within a projected distance of 1.0~Mpc and a line-of-sight velocity interval of $\Delta v_r = 500$ km/s.

On the other hand, collisions of galaxies would result in gas stripping and quenching of the star formation, which has been shown in \cite{Benavides21}, where a sample of isolated Ultra-Diffuse Galaxies (UDGs) in the TNG50 simulation was analyzed. UDGs are not dwarfs, their dynamical masses may be quite large, but they have low stellar masses, comparable to those of dwarfs. The authors of \cite{Benavides21} discovered  that isolated UDGs in the simulation are actually backsplash galaxies. These galaxies can be located over 1~Mpc away from their targets and can have velocities of more than 500 km/s relative to their local environment. In general, backsplashers move away from their targets and have significant velocities with respect to their environment, with median about 200~km/s \cite{Benavides21,Osipova23}. 

 Following the methodology of \cite{Benavides21}, a recent analysis focused on quenched field dwarfs \cite{Benavides25} revealed that gas stripping by filaments may also be an important process for isolated quiescent dwarf galaxies. Besides, there is observational evidence supporting the presence of a backsplash candidate with quenched star formation \cite{Casey22}.

The comparison of studies \cite{Benavides21,Osipova23,Benavides25} and \cite{Bidaran_2025} points to the fact that the isolation criterion based on the distance of 1~Mpc and velocity distance $\Delta v_r < 500$ km/s may be insufficient for reliable identification of galaxies that had no flyby events in the past. Then, the question arises: what is the maximal velocity a backsplash galaxy could reach in principle? Although simulations can provide insights into this question, they come with certain limitations. In particular, when studying dwarf galaxy haloes, high resolution is required, which consequently limits the ability to simulate large
cosmological volumes of real surveys.

In this paper we develop a simple theory which can predict a maximal velocity of a backsplash galaxy based on the well known Zel'dovich approximation \cite{Zeldovich70}. We test this theory by applying it to a sample of backsplash galaxies identified in a high resolution cosmological simulation.

The paper is organized as follows: in Section~\ref{sec:model} we introduce the model of galaxy collisions. In Section~\ref{sec:vmax} this model is used in conjunction with the Extreme Value Statistics to derive the maximal collision velocity and maximal backsplash velocity in a given volume. In Section~\ref{sec:concl} we summarize our findings and discuss their applications. In the Appendix~\ref{sec:app} we provide details on the calculation of the distribution function of collision velocities.
 
\section{Model of galaxy collisions}\label{sec:model}
In this Section we introduce the model, and also test some of its assumptions using a cosmological simulation of dark matter haloes called Extremely Small MultiDark Planck (ESMDPL) carried out by an international consortium within the framework of the MultiDark project\footnote{\url{http://www.multidark.es}} using the LRZ supercomputing facility. It has a box size of 64~Mpc/h filled with $4096^3$ dark matter particles. The simulation data is available in CosmoSim\footnote{\url{https://www.cosmosim.org}} database. 

There is a degree of uncertainty around the terminology in the area of haloes passing through another halo: backsplash haloes are often called ``flyby''. Moreover, the definitions of these haloes vary in the literature \cite{Gill:Knebe:Gibson:2005, Teyssier:Johnston:Kuhlen:2012, Diemer_2021, Haggar:Gray:Pearce:2020, Mansfield:Kravtsov:2020}. We use the sample of backsplash haloes similar to that found in \cite{Osipova23}. These are field haloes that had only one collision in the past by our definition, which differs from some other works where objects with several collisions are also called backsplash haloes, e.g. \cite{Diemer_2021}. We trace the main progenitors of both the backsplash haloes and their targets before the collisions using merger trees built with the Consistent Trees code \cite{Behroozi:Wechsler:Wu:2013:763:18}. 

\hfill \break
In this paper, to be called backsplash, a halo must satisfy the following conditions:
\begin{flushleft} 
\begin{enumerate}
    \item a backsplash halo is a field halo at $z = 0$, 
    \item its main progenitor was within the virial radius of another more massive halo, the target, only once during the time when the halo finder is able to track its evolution (this time is different for each halo),
    \item a backsplash halo and the descendant of the target are different haloes at $z = 0$,
    \item target's main progenitor was a field halo during the time when the halo finder is able to track its evolution.
    
\end{enumerate}
\end{flushleft}
\hfill \break
In order to identify backsplash haloes, as well as their
targets, we use the passage detection algorithm described in \cite{Osipova23}. Taking into account our earlier research, which revealed that many backsplash haloes with less than $\sim 100$ particles within the virial radius are not identified correctly, low-massive haloes should be used with caution. Therefore, we examine only backsplash haloes that contain more than 24 particles at the moment of the passage through another halo and have more than 100 particles within their virial radius at z = 0.

In our model we assume that progenitors of backsplash galaxies and their targets move in accordance with the Zel'dovich approximation long before the collision.
One can write the distance between two galaxies $d$ and their relative velocity $v$ in projection to the line connecting two galaxies:
\begin{align}
\label{eq:za_delta}
    d &= \Delta s + D(a) \Delta \chi, \\
    \label{eq:za_delta_v}
    v &= H(a)d + a\dot{D}\Delta \chi,
\end{align}
where $\Delta s$ is the Lagrangian distance, $\Delta \chi$ is the difference of the displacements, projected to the line connecting two galaxies, $a$ is the scale factor, $H(a)$ is the Hubble constant, and $D$ is the linear theory growth factor. The dot marks differentiation by time.
According to this Zel'dovich approximation, collision happens when $d=0$, or:
\begin{align}
\label{eq:za_collision_a}
    \Delta s &= - D(a_c) \Delta \chi, \\
    \label{eq:za_collision_v}
    v_c &= a_c\dot{D}(a_c)\Delta \chi.
\end{align}
If $\Delta s$ and $\Delta \chi$ are known, equation (\ref{eq:za_collision_a}) can be used to find the scale factor of the collision, $a_c$, and (\ref{eq:za_collision_v}) gives the velocity. Also equation (\ref{eq:za_collision_a}) implies that collision could happen only if $|\Delta \chi| \ge |\Delta s|$, since $D(a)\le 1$ for $a\le 1$.

Equations (\ref{eq:za_delta})-(\ref{eq:za_collision_v}) describe the collision of two massles test particles. For real galaxies, however, their motion will differ from equations (\ref{eq:za_delta})-(\ref{eq:za_collision_v}) when they become sufficiently close to each other. We assume that equations (\ref{eq:za_delta})-(\ref{eq:za_delta_v}) could be applied when
\begin{equation}
\label{eq:za_acc_req}
    |\dot{v}| > \frac{G(M+m)}{d^2}.
\end{equation}
Here  $M$ and $m$ are the masses of the target and the backsplash progenitors, respectively. We test this assumption using the sample of backsplash progenitors in the ESMDPL simulation. For that we first measure $d_1$ and $v_1$ at some early moment of time with scale factor $a_1$,which allows us to determine $\Delta s$ and $\Delta \chi$. Using these values, we compute $d_z$ and $v_z$ at $a_2>a_1$ and compare them with measured distances $d_2$ and velocities $v_2$ if the condition (\ref{eq:za_acc_req}) is satisfied.

\begin{figure*}
    \centering
    \includegraphics[width=0.98\textwidth]{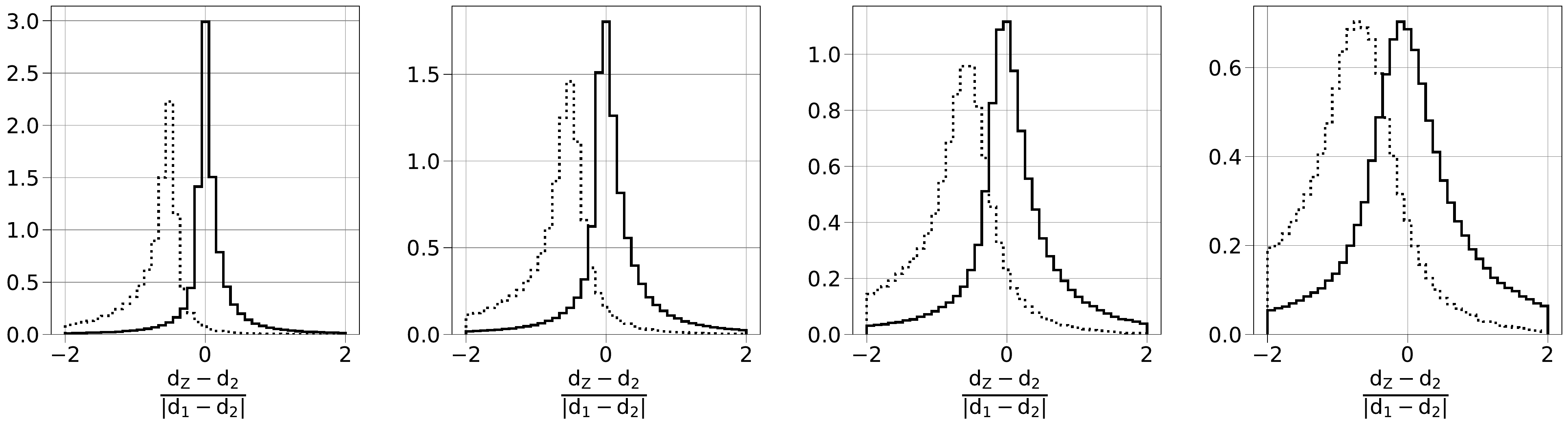}
    \caption{ Distributions of relative deviations in the distances between backsplashers and their targets, comparing predictions from the Zel’dovich approximation, $d_z$, with simulation measurements, $d_2$. The difference between the predictions and actual simulation distances is divided by the change in backsplash and target separation during observation time, $|d_1-d_2|$. The distances were compared at the time of the last snapshot before the collision, where the Zel’dovich acceleration does not exceed the internal acceleration of the system, defined as the effective mass acceleration in the two-body problem (\ref{eq:za_acc_req}). The dashed line represents the distribution of a similarly computed errors for inertial motion. The distributions are shown for different scale factor intervals of $a_2/a_1=1.1$, 1.2, 1.4, 1.8 (solid lines, from left to right).}
    \label{fig:ZA_rel_d_err}
\end{figure*}

We plot the relative errors of distance prediction, $(d_z-d_2)/(d_1-d_2)$ in Fig.~\ref{fig:ZA_rel_d_err} for different ratios of $a_2/a_1$. As one can see from this Figure, the distribution of errors peaks around zero, which means that Zel'dovich approximation reproduces the motion of galaxies before the collision quite well. As an illustration, in Fig.~\ref{fig:ZA_rel_d_err} are also shown predictions of inertial motion, i.e. ignoring the accelerations created by the large scale structure.

\begin{figure*}
    \centering
    \includegraphics[width=0.49\textwidth]{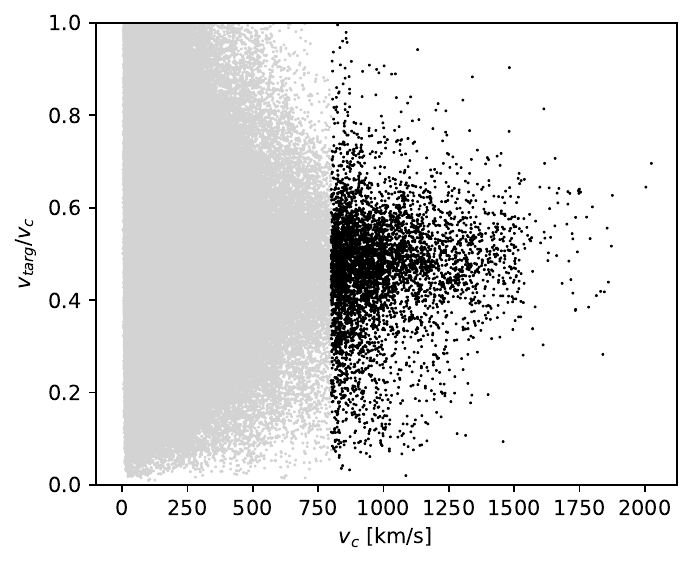}
    \includegraphics[width=0.49\textwidth]{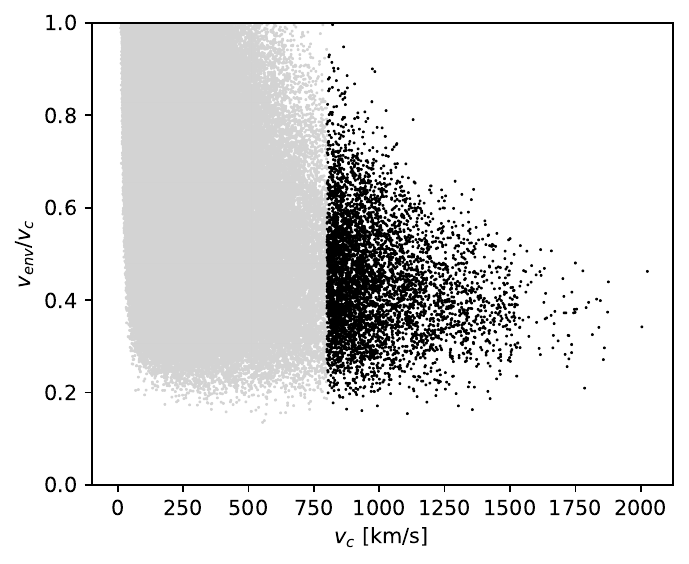}
    \caption{Ratio of the backsplash velocity with respect to its target $v_{targ}$ (\textit{left panel}), or with respect to its environment $v_{env}$ (\textit{right panel}) to the predicted collision velocity $v_c$. Darker dots show systems with $v_c>800$~km/s.}
    \label{fig:v_corr}
\end{figure*}

Our next assumption is that the system of two bodies, the backsplash and the target, retain some memory of the initial velocity after the collision. Indeed, for a very simplified model of two point masses on a hyperbolic orbit this is the case. For real galaxies, we expect a correlation between the collision velocity predicted by the Zel'dovich approximation (\ref{eq:za_collision_v}) and the present-day velocity of backsplash haloes with respect to their targets, $v_{targ}$, as well as with respect to the environment, $v_{env}$.

We perform an assessment of this correlation using the ESMDPL simulation. Our analysis reveals that in 95\% of systems, the condition $|\Delta \chi| \ge |\Delta s|$ is met, thereby enabling the use of equation (\ref{eq:za_collision_a}) to find the moment of collision. The results are shown in Fig.~\ref{fig:v_corr}. We are now interested only in the highest velocity backsplash haloes, since our aim is to predict the absolute maximum. Thus, we restrict our analysis to systems with a critical velocity $v_c>800$~km/s, which is approximately half of the maximal value of $v_c$ in our sample. The results demonstrate a relatively weak sensitivity to changes in this boundary when moved to higher velocities. For that subsample, shown with darker dots in Fig.~\ref{fig:v_corr} we find that
\begin{equation}
\label{eq:corr-targ}
    v_{targ} = (0.47\pm 0.13) v_c,
\end{equation}
\begin{equation}
\label{eq:corr-env}
    v_{env} = (0.45\pm 0.12) v_c,
\end{equation}
where the root mean square of the deviation from the mean is shown after the `$\pm$' sign. We define velocities with respect to the surrounding field haloes, $v_{env}$, using the same approach as in \cite{Benavides21}: we select field haloes within 1~$h^{-1}\; \mathrm{Mpc}$ around every backsplash halo and calculate the average absolute value of the peculiar velocity difference between the neighbour and the backsplash halo.

Now we can use correlations (\ref{eq:corr-targ}) and (\ref{eq:corr-env}) to predict velocities of backsplash galaxies using statistical properties of the displacement field and Zel'dovich approximation.

\section{Maximal velocity of a backsplash}\label{sec:vmax}
From the analysis of equations (\ref{eq:za_collision_a})-(\ref{eq:za_collision_v}), it can be seen that for a given Lagrangian distance $\Delta s$, there is an upper bound on the collision velocity, $v_{c,lim}$. This limitation arises because when $|\Delta \chi|$ is large, the collision happens earlier, resulting in a lower value of $a_c\dot{D}(a_c)$.
Therefore, the collision velocities range from $0$ to $v_{c,lim}$. The probability distribution function $f(v_c)$ depends on the distribution of $\Delta \chi$  and is derived in Appendix~\ref{sec:app}.

Now we can use the Extreme Value Statistics (EVS) to find the maximal expected $v_c$. For a sequence of independent measurements of length $N$ derived for a value following the probability density function (PDF) $f(x)$ the cumulative distribution function (CDF) of the maximal value $x_{max}$:
\begin{equation}
\label{eq:EVS}
    F_{max}(x_{max}) = [F(x_{max})]^{N},
\end{equation}
where $F(x)=\int_{-\infty}^x f(y) dy$ is the cumulative distribution function, corresponding to $f$.

\begin{figure}
    \centering
    \includegraphics[width=1.0\linewidth]{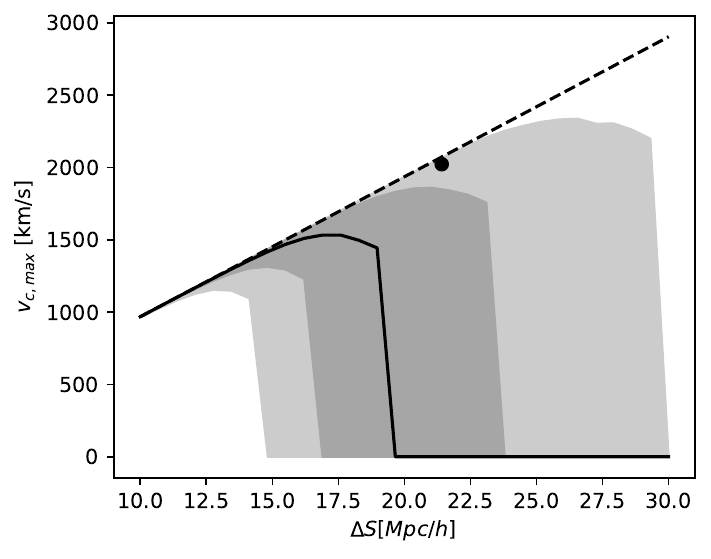}
    \caption{Distribution of the maximal collision velocity in the (64~Mpc/h)$^3$ volume: the solid line is the median, the darker shaded region corresponds to 1 standard deviation, the lighter shaded region represents 2 standard deviation. The dot corresponds the maximal value from the ESMDPL simulation. The dashed line is the maximal possible collision velocity $v_{c,lim}$.}
    \label{fig:vcmax}
\end{figure}

For a fixed Lagrangian distance $\Delta s$ we expect the measurements are independent if the systems are located on the distance $>\Delta s$ from each other. Thus, for a given volume $V$ we can estimate 
\begin{equation}
\label{eq:N}
    N=V/\Delta s^3.
\end{equation}

There may be a coefficient $\sim1$ before $\Delta s^3$ in equation (\ref{eq:N}), however,  its influence is negligible, since $F_{max}$ rather weakly depends on $N$. Applying equations (\ref{eq:EVS}), (\ref{eq:N}) to the CDF of $v_c$ obtained in the Appendix~\ref{sec:app}, we find the CDF of maximal $v_{c,max}$ for a given $\Delta s$ in a given volume, from which we calculate the median as well as $1\sigma$ and $2\sigma$ deviations.

The median value and the regions with probability corresponding to $1\sigma$ and $2\sigma$ deviations from the median for a simulation with 64~Mpc/h box size are shown in Fig.~\ref{fig:vcmax}. 
The observed behavior of the maximal velocity distribution illustrated in Fig.~\ref{fig:vcmax} can be explained as follows. For small $\Delta s$, there are many realizations of colliding haloes in the given volume and thus the maximal collision velocity is very close to the limit $v_{c,lim}$ imposed by equations (\ref{eq:za_collision_a})-(\ref{eq:za_collision_v}). For sufficiently large values of $\Delta s$, we approach the homogeneity scale of the Universe, where the small amplitude of density perturbations prevents collisions from occurring. Therefore, $v_{c,max}$ tends toward zero as $\Delta s$ becomes large.
The calculation of  $v_{c,max}$ for a range of $\Delta s$ values enables us to ascertain the absolute maximum collision velocity within a given volume. 
By finding the peak of the median as a function of $\Delta s$ we obtain the value, such that half of the random realizations of the displacement field have maximal velocities below this point. If we shift our focus from the median to the $P=84\%$, $P=16\%$ and $P=97.7\%$, $P=0.023\%$ percentiles, we obtain the intervals of the maximal value corresponding $1\sigma$ or $2\sigma$ deviation.

Fig.~\ref{fig:vcmax} demonstrates that the simulation result falls within the $2\sigma$ region of our prediction.  It should be emphasized that ESMDPL is a constrained simulation, so the amplitudes of its large scale modes were to create an accurate local environment despite the small box size. Consequently, this adjustment might result in the relatively high value of the maximal velocity compared to the median of our predictions.

\begin{figure}
    \centering
    \includegraphics[width=\linewidth]{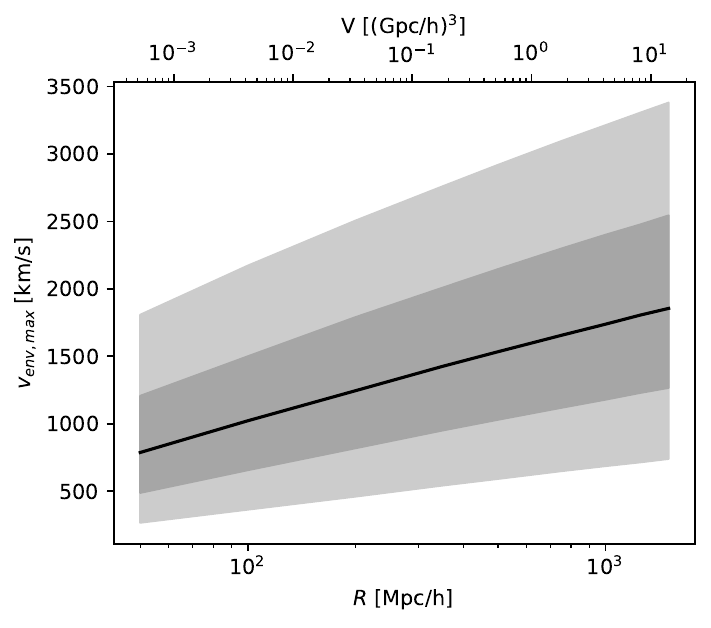}
    \caption{Expected maximal velocity of a backsplash in a sphere of radius $R$ (or volume $V$) relative to the environment, $v_{env}$: the median (solid line) and regions with probability $P=16-84\%$ (dark grey) and $P=0.023-97.7\%$ (light grey).}
    \label{fig:vmax-vol}
\end{figure}

Combining the EVS results with the correlations (\ref{eq:corr-targ}) and (\ref{eq:corr-env}) we finally find the theoretical estimates for the maximal values in the the ESMDPL simulation box with three probabilities, $P=50\%$ (median), $P=84\%$ ($1\sigma$) and $P=97.7\%$ ($2\sigma$), respectively:
\begin{equation}
    v_{targ} \approx v_{env} = 700,\;\;1100,\;\;1700 \;\text{km/s}
\end{equation}
while the actual maximal velocities are:
\begin{align}
    &v_\text{targ,ESMDPL} = 1400\;\text{km/s}, \\
    &v_\text{env,ESMDPL} = 940\;\text{km/s}.
\end{align}
It is seen that these values are within the $1-2\sigma$ interval of the theoretical estimate.
The estimates for several volumes in the interesting for observations range are given in Figure~\ref{fig:vmax-vol}. The scale of perturbations, $\Delta s$, which produces the highest velocities in Figure~\ref{fig:vmax-vol}, ranges from 20 to 50~Mpc/h.

\section{Conclusions}\label{sec:concl}
We have developed a method that allows us to predict the maximal velocity of a backsplash galaxy relative to its environment. The method is described in Sections~\ref{sec:model} and~\ref{sec:vmax}, and the main result is presented in Figure~\ref{fig:vmax-vol}. Several implications can be drawn from these results. Based on our theoretical estimates and also direct analysis of the ESMDPL simulation, which has a box size of 64~Mpc/h, we conclude that backsplash galaxies should exist with velocities of $\approx 1000$~km/s relative to their environment in the Local Universe within a  few tens of Mpc from us. The same conclusion should hold for every patch of the Universe of comparable size. For larger volumes, extending out to 1500 Mpc/h from us, this maximal speed can exceed 2000~km/s.
Similar maximal velocities are expected between a backsplash galaxy and its target.

These results are  critical for identifying quiescent isolated galaxies, which are used to test models of galaxy quenching. If the distances to analyzed galaxies are determined from their redshifts, the enormous velocities of some rare backsplashers can lead to errors exceeding 10~Mpc. Such errors could completely change the apparent environment of a galaxy. Additionally, backsplashers can travel large distances from their past targets during the cosmic time: a velocity of $1000$~km/s corresponds to approximately 1~Mpc/Gyr. As a result, a fast backsplasher could be found several Mpc away from the galaxy with which it interacted in the past. For example in the ESMDPL simulation the largest observed separation between a backsplash halo and its target is 2.9~Mpc/h. Backsplash galaxies, having been processed during their collisions, are expected to be quiescent. Therefore, their presence in samples of `isolated' galaxies could compromise the results of quenching studies.

Considering the data presented in Figure~\ref{fig:vmax-vol}, we suggest raising the minimal allowed projected distance and velocity difference in the isolation criterion beyond 1~Mpc and 500~km/s. However, this approach would likely result in a substantial decrease in the number of isolated objects, potentially resulting in none remaining. Another approach to improve the analysis of isolated quiescent galaxies is to predict the expected number of backsplashers for a given survey and isolation criterion. This can be achieved using mock catalogues generated from simulations, or, potentially, by further developing the theoretical framework introduced in this paper.

Our approach is also applicable to the recent findings of \cite{Benavides25} who discovered that dwarfs can become quenched as a result of their passages not only through other galaxies, but also through filaments. Although the calculation of $v_{c,max}$ remains unchanged in that case, the ratio between the $v_c$ and $v_{env}$ may differ from what we derived in equation (\ref{eq:corr-env}).

Our results can be safely applied to haloes with masses below few $10^{12}$~M$_\odot$, as discussed in the Appendix~\ref{sec:app}. For higher masses the theory needs to consider the constraints on the displacement field arising from the fact that the haloes themselves have already formed. Consequently, our calculations cannot be applied to objects such as the renowned Bullet cluster.

\begin{acknowledgments}
The authors gratefully acknowledge the Gauss Centre for Supercomputing e.V. (www.gauss-centre.eu) for funding this project by providing computing time on the GCS Supercomputer SUPERMUC-NG at Leibniz Supercomputing Centre (\url{www.lrz.de}). We thank Peter Behroozi for creating and providing the ROCKSTAR merger trees of the ESMDPL simulation. The CosmoSim database (\url{https://www.cosmosim.org}) provides access to the simulation and the ROCKSTAR data. The database is a service by the Leibniz Institute for Astrophysics Potsdam (AIP).
\end{acknowledgments}

\appendix

\section{Distribution function of the collision velocity}\label{sec:app}
The displacement field $\vec{\chi}(\vec{s})$ is Gaussian and its properties can be obtained from the cosmological power spectrum of density perturbations $P(k)$. Let's consider two points with Lagrangian coordinates $\vec{s}_1$ and $\vec{s}_2$, the difference of displacements in these points, $\Delta \vec{\chi} = \vec{\chi}(\vec{s}_2)-\vec{\chi}(\vec{s}_1)$. As has been shown in, e.g., \cite{DD99}
\begin{equation}
\label{eq:sigma_chi}
    \sigma^2_{\chi}(\Delta s) \equiv \langle \Delta \vec{\chi}^2 \rangle = \frac{1}{\pi^2} \int_0^\infty P(k)\left( 1 - \frac{\sin(k \Delta s)}{k\Delta s}  \right) dk.
\end{equation}

\begin{figure}
    \centering
    \includegraphics[width=0.99\linewidth]{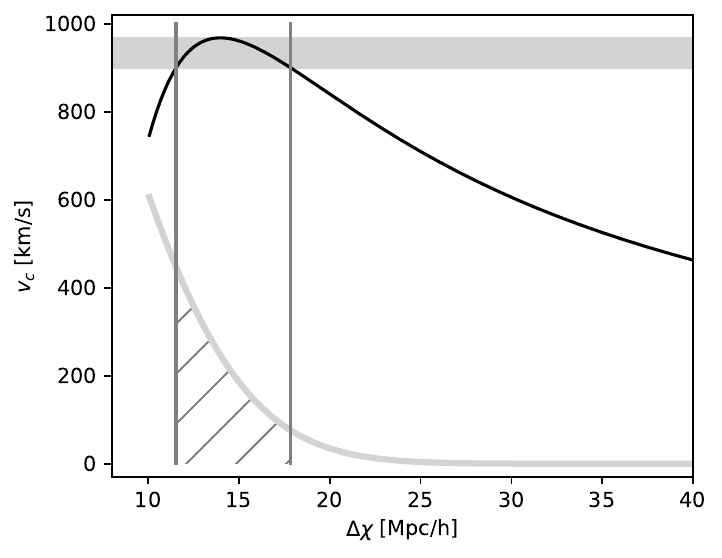}
    \caption{A sketch representing the relation between distribution functions of $\Delta \chi$ and $v_c$. The black solid line shows the relation between $\Delta \chi$ and $v_c$. A region is shaded between some arbitrary chosen $v_c=900$~km/s and $v_{c,lim}$. The gray thick line is the PDF of $\Delta \chi$ amplified to fit the plot range. Vertical lines show values of $\Delta \chi$ where $v_c=900$~km/s. The hatched region shows the integral of $\Delta \chi$ PDF between the vertical lines, and it equals to the probability to find $v_c$ between 900~km/s and $v_{c,lim}$.}
    \label{fig:sketch}
\end{figure}

Considering that the backsplash progenitor is expected to collide with the target progenitor, we can assume that $\Delta\vec{s}$ and $\Delta\vec{\chi}$ are anti-parallel due to our choice of $\vec{s}_1$ and $\vec{s}_2$. This allows us to determine the direction of $\Delta\vec{\chi}$, and equation (\ref{eq:sigma_chi}) now provides the dispersion of the projection of $\Delta \vec{\chi}$ onto the line connecting the two galaxies. Thus, the value of $\Delta \chi$ in (\ref{eq:za_delta})-(\ref{eq:za_collision_v}) follows a Gaussian distribution with zero mean and dispersion given by equation (\ref{eq:sigma_chi}). The dispersion $\sigma_\chi$ increases with $\Delta s$, gradually approaching its maximum value of $\sigma_\chi(\Delta s \rightarrow \infty)=13.4$~Mpc/h.

These considerations on the distribution of $\Delta \chi$ are obtained for massless test particles. A dark matter halo of mass $m$ is assembled from a Lagrangian region of size
\begin{equation}
    \label{eq:lagr_size}
    s_{m}\sim \left( \frac{m}{\bar{\rho}} \right)^{1/3},
\end{equation}
where $\bar{\rho}$ is the mean matter density of the Universe. Strictly speaking, in order to compute the distribution of the relative displacements of pairs of haloes with finite masses $m$, one should first smooth the displacement field on the scale $s_m$ and then compute the probability distribution assuming that the Lagrangian regions of both haloes have already collapsed to form these haloes. Similar calculations have been performed in \cite{DD99p2} for 1D pancakes. While extending this to  3D haloes  would involve a much more complicated calculation, we argue that small haloes with $s_m\ll \Delta s$ can be effectively treated as massless particles, making equation (\ref{eq:sigma_chi}) valid for them. As we will see later, the maximal collision velocities are determined by displacements on scales $\Delta s =20-50$~Mpc/h, and a halo with mass $10^{12}$~M$_\odot$/h has Lagrangian size of only $s_m\approx2$~Mpc/h. Smoothing the displacement field with the top-hat filter on the scale $s_m$ does not significantly alter the dispersion $\sigma_\chi$ on the scale of 20~Mpc/h. Also, since the displacement field is Gaussian, the Fourier modes of this field with different wave numbers are independent. Therefore, the conditions imposed on the properties of haloes' Lagrangian regions should not impact the probability distribution of displacements on much larger scales. 

We need to transform the Gaussian distribution of $\Delta \chi$ into the distribution of $v_c$. The connection between $v_c$ and $\Delta \chi$ is given by equations (\ref{eq:za_collision_a})-(\ref{eq:za_collision_v}), and an example for $|\Delta s| = 10$~Mpc/h is shown in Fig.~\ref{fig:sketch} by the dark line. Function $v_c(\Delta \chi)$ is a non-monotonic one, it has one maximum, $v_{c,lim}$. Since the fact of collision requires $|\Delta \chi| \geq |\Delta s|$, the equation  $v_c(\Delta \chi) = v $ has two solutions when $v \ge v_c(|\Delta s|)$, and one solution when $v < v_c(|\Delta s|)$. When there are two solutions $\Delta \chi_1$, $\Delta \chi_2$, the CDF of $v_c$ is given by:
\begin{equation}
    F_{v_c}(v) = 1 - (F_\chi(\Delta \chi_2)-F_\chi(\Delta \chi_1)),
    \label{eq:Fvc1}
\end{equation}
where $F_\chi$ is the CDF of $\Delta \chi$ (i.e. the Gaussian distribution).
The origin of this equation is illustrated by the sketch in Fig.~\ref{fig:sketch}. When there is only one solution $\Delta \chi_2$, the connection changes to:
\begin{equation}
    F_{v_c}(v) = 1 - (F_\chi(\Delta \chi_2) - F_\chi(|\Delta s|)).
    \label{eq:Fvc2}
\end{equation}
The PDF $f_{v_c}$  now can be obtained by differentiating (\ref{eq:Fvc1})-(\ref{eq:Fvc2}).


 \bibliographystyle{elsarticle-num} 
 \bibliography{vel}





\end{document}